\documentclass[aps,prd,twocolumn,superscriptaddress,preprintnumbers,floatfix,nofootinbib,notitlepage,showkeys]{revtex4-1}

\usepackage[utf8x]{inputenc}
\usepackage{cancel}

\usepackage{graphicx}
\usepackage{hyperref}
\usepackage{latexsym}
\usepackage{amsmath}
\usepackage{amssymb}
\usepackage{bbm}

\usepackage{ulem}
\usepackage{pdfsync}
\usepackage{epsfig}
\usepackage{epstopdf}
\usepackage{subfigure}
\usepackage{color}
\usepackage{comment}
\usepackage{slashed}

\def\beq{\begin{equation}}
\def\eeq{\end{equation}}
\def\baq{\begin{eqnarray}}
\def\eaq{\end{eqnarray}}

\newcommand{\be}{\begin{equation}} 
\newcommand{\ee}{\end{equation}}
\newcommand{\bea}{\begin{eqnarray}} 
\newcommand{\eea}{\end{eqnarray}}

\newcommand{\bmp}{\noindent\begin{minipage}{16cm}}
\newcommand{\emp}{\end{minipage}\vskip 7mm} 
\def\lsim{\mathrel{\raise.3ex\hbox{$<$\kern-.75em\lower1ex\hbox{$\sim$}}}}
\def\gsim{\mathrel{\raise.3ex\hbox{$>$\kern-.75em\lower1ex\hbox{$\sim$}}}}

\newcommand{\intron}[1]{}

\begin{document}

\title{Prospects for indirect detection of frozen-in dark matter}

\author{Matti Heikinheimo}
\email{matti.heikinheimo@helsinki.fi}
\affiliation{Department of Physics, University of Helsinki \\
                      P.O.~Box 64, FI-00014, Helsinki, Finland}
\affiliation{Helsinki Institute of Physics, \\
                      P.O.~Box 64, FI-00014, Helsinki, Finland}
\author{Tommi Tenkanen}
\email{t.tenkanen@qmul.ac.uk}
\affiliation{Astronomy Unit, Queen Mary University of London, \\
                      Mile End Road, London, E1 4NS, United Kingdom}
\author{Kimmo Tuominen}
\email{kimmo.i.tuominen@helsinki.fi}
\affiliation{Department of Physics, University of Helsinki \\
                      P.O.~Box 64, FI-00014, Helsinki, Finland}
\affiliation{Helsinki Institute of Physics, \\
                      P.O.~Box 64, FI-00014, Helsinki, Finland}

\begin{abstract}
We study observational consequences arising from dark matter (DM)
of non-thermal origin,
produced by {\it dark freeze-out} from a hidden sector heat bath. We assume this heat bath was
populated by feebly-coupled mediator particles, produced via a Higgs portal interaction with the Standard Model (SM).
The dark sector then attained internal equilibrium with a characteristic temperature
different from the SM photon temperature.
We find that even if the coupling between the DM and the SM sectors is very weak, the scenario allows for indirect observational signals. We show how the expected strength of these signals depends on the temperature of the hidden sector at DM freeze-out.
\end{abstract}

\preprint{HIP-2018-1/TH}

%
\maketitle

%
\section{Introduction}

There is overwhelming evidence for a long-lived, relatively cold, non-baryonic
matter component, whose abundance in the Universe exceeds the amount of
ordinary matter roughly by a factor of five, and which has been there from the
hot Big Bang era until the present day~\cite{Ade:2015xua}. While the existence
of Dark Matter (DM) seems indisputable, its non-gravitational nature remains a
mystery~\cite{Bergstrom:2000pn,Arcadi:2017kky}.

In the standard Weakly Interacting Massive Particle (WIMP) paradigm,
DM is assumed to interact with the visible sector strongly enough to
have been initially in thermal equilibrium with it. At some point
the expansion rate of the Universe overcame the interaction rate between
DM and the visible sector leading to freeze-out of the DM relic density.
An appealing aspect of
this scenario is that the same interaction that determines the DM abundance is
also responsible for making the paradigm testable by terrestrial experiments,
although so far they have yielded only null results~\cite{Arcadi:2017kky}. For
a DM particle mass not too different from the electroweak scale, this forces
the coupling between DM and the visible sector to be small.

If the coupling between DM and the visible sector was very small, the DM particles were never in thermal equilibrium with the SM particles. In that case, the DM abundance has to be produced non-thermally, for example by the so-called {\it freeze-in} mechanism instead of the usual freeze-out paradigm \cite{McDonald:2001vt,Hall:2009bx}. The freeze-in production typically requires very small couplings, $\lambda\lesssim 10^{-7}$, and the corresponding DM particle is called a Feebly Interacting Massive Particle (FIMP). In this scenario, the DM particles are produced by decays and annihilations from the visible sector, until the production ceases due to the cooling of the photon temperature below the relevant mass scale connecting the DM particle to the visible sector. For a recent review of freeze-in scenarios considered in the literature, see \cite{Bernal:2017kxu}.

The weakness of interactions between the DM and the SM particles in the freeze-in scenario implies that these models are inherently very difficult to search for in direct detection or collider experiments.
However, this is turning into an appealing feature as the experimental constraints are beginning to rule out large parts of the parameter space of the typical WIMP models.
On the other hand, possibilities for observing feebly-coupled DM indirectly exist.
One such possibility
would be to allow for a non-vanishing mixing angle between a singlet DM fermion
and the SM neutrinos, which has well-known observable consequences  \cite{Adulpravitchai:2014xna,Merle:2014xpa,Merle:2015oja,Shakya:2015xnx, Heikinheimo:2016yds}.
Other possibilities to probe non-thermal, frozen-in DM include studies of
formation of small scale structure of the Universe \cite{Konig:2016dzg,Murgia:2017lwo} or imprints on the Cosmic Microwave
Background radiation \cite{Nurmi:2015ema,Kainulainen:2016vzv,Heikinheimo:2016yds,Tenkanen:2016twd,Enqvist:2017kzh}.
Another possibility, which we will explore in this paper, is to allow for
DM annihilations into mediator particles that will eventually decay into SM particles.

An example of a signature that would result from the possibility that DM annihilates into unstable mediator species is the claimed detection of a galactic center excess at a GeV energy scale \cite{Goodenough:2009gk,Hooper:2010mq,Daylan:2014rsa,Calore:2014xka,Escudero:2017yia}. In the case where frozen-in, stable DM particles $A$ annihilate into unstable mediator particles $s$, which then decay into the SM, the cross section of the process $AA\rightarrow ss$ does not have to be very suppressed (as compared to the case where the DM abundance was generated by the freeze-out mechanism), while the hidden sector remains secluded from the SM due to small couplings between the mediator $s$ and the SM. Such mechanism has previously been applied in the context of the galactic center gamma ray excess in~\cite{Heikinheimo:2014xza,Biswas:2015bca,Escudero:2017yia}.

In this paper, we investigate a benchmark scenario where the DM particle $A$ is a spin-1 vector boson of a hidden gauge symmetry, 
and which was never in thermal equilibrium with the SM. In our case, in contrast to~\cite{Heikinheimo:2014xza,Biswas:2015bca,Escudero:2017yia}, the temperature
of the hidden sector differs from that of the SM. We will demonstrate how this
is reflected on the magnitude of the relevant
annihilation cross section, both for the production of the correct DM relic density and the indirect detection signal:
If the hidden sector temperature during the dark freeze-out process is smaller than the SM temperature, the corresponding equilibrium yield of DM particles is smaller and thus the dark freeze-out must happen earlier in order to produce the observed abundance. Therefore the annihilation cross section that results in the correct DM abundance is smaller than in the case of equal hidden and visible sector temperatures, and consequently the expected indirect detection signal is weaker.

Our results are applicable beyond the simple model setup we consider.
Similar features are expected to emerge in many FIMP scenarios where the
DM abundance is determined by freeze-out of an s-wave annihilation process
to lighter mediator particles within a hidden sector.

The paper is organized as follows: in Sec. \ref{model} we present the model under investigation and then study the DM production in Sec. \ref{abundance}. In Sec. \ref{indirect_detection}, we discuss the observational prospects within this model class related to indirect detection. Finally, in Sec. \ref{conclusions} we present our conclusions.

\section{The Model}
\label{model}

We consider a scenario where the DM particle is a \mbox{spin-1} vector boson. As representative examples we study the hidden vector DM model of \cite{Hambye:2008bq}, where the DM is a triplet of massive gauge bosons of a broken $SU(2)$ gauge group, and a simpler model, where the DM is a massive gauge boson of a broken $U(1)$ gauge group. In both scenarios, in addition to the DM particle there is a scalar $s$ acting as a mediator between the hidden sector and the SM. The scalar $s$ is a complex doublet of a hidden $SU(2)_{\rm D}$ gauge symmetry, or charged under the hidden $U(1)$ gauge symmetry, respectively in the two scenarios, but singlet under the SM gauge groups. Thus, the hidden sector interaction Lagrangian is
\begin{equation}
\mathcal{L}_{\rm hidden} =  \frac{1}{4}F^{'\mu\nu}F^{'}_{\mu\nu} + (D^{\mu}s)^{\dagger}(D_{\mu}s) ,
\end{equation}
where $D_{\mu}s = \partial_{\mu}s -i\frac{g'}{2}\tau^a A^a_\mu s$ with $A^a_\mu$ the $SU(2)_{\rm D}$ gauge fields, in the case of $SU(2)$, and $D_{\mu}s = \partial_{\mu}s -\frac{i}{2} g' A_\mu s$ in the case of $U(1)$, $\tau^a$ are the Pauli matrices and $F^\prime$ the field strength associated with the gauge field $A_\mu$. The scalar potential we take to be
\begin{eqnarray}
\label{potential}
V(\Phi,s) = &-&\mu_{\rm h}^2\Phi^\dagger\Phi
+\lambda_{\rm h}(\Phi^\dagger\Phi)^2
- \mu_{\rm s}^2 s^{\dagger}s \\ \nonumber
&+& \lambda_{\rm s}(s^{\dagger}s)^2 + \lambda_{\rm hs}\Phi^\dagger\Phi s^{\dagger}s ,
\end{eqnarray}
with standard kinetic terms. Here $\Phi$ is the SM Higgs doublet, which obtains a vacuum expectation value (vev), $\sqrt{2}\Phi = (0,v+h)$,
where $v=246$ GeV. We assume the same normalization for the $s$ field.

The stability of the scalar potential requires $\lambda_{\rm hs} > -2\sqrt{\lambda_{\rm h}\lambda_{\rm s}}$ and $\lambda_{\rm s}>0$. We assume $\mu_{\rm s}^2<0$ in order to induce a vev for the singlet scalar $s$ and spontaneously break the hidden sector gauge symmetry\footnote{A scenario where the hidden sector exhibits a scale invariance which is spontaneously broken through the portal coupling $\lambda_{\rm hs}$ in the electroweak (EW) phase transition was studied in~\cite{Heikinheimo:2017ofk}. In that case, the masses
are small and subtly related to the EW scale $v$. However, here we allow a more general setting to study the observational consequences also at the GeV energy range.}, which we take to be the sole mechanism to generate a mass for the DM particle $A$.

Thus, after the SM Higgs and the singlet gain vevs
\bea
v &=& \sqrt{2}\frac{\sqrt{2\lambda_{\rm s}\mu_{\rm h}^2-\lambda_{\rm hs}\mu_{\rm s}^2}}{\sqrt{4\lambda_{\rm h}\lambda_{\rm s}-\lambda_{\rm hs}^2}}\approx \frac{\mu_{\rm h}}{\sqrt{\lambda_{\rm h}}}=246\ {\rm GeV}, \\ \nonumber
v_{\rm s} &=& \sqrt{2}\frac{\sqrt{2\lambda_{\rm h}\mu_{\rm s}^2-\lambda_{\rm hs}\mu_{\rm h}^2}}{\sqrt{4\lambda_{\rm h}\lambda_{\rm s}-\lambda_{\rm hs}^2}} \approx \frac{\mu_{\rm s}}{\sqrt{\lambda_{\rm s}}},
\eea
the vector boson mass is given by
\be
m_{\rm A} = \frac{1}{2}g' v_s.
\ee
The scalars $s$ and $h$ mix due to the mass matrix
\be
\mathcal{M}^2 = \begin{pmatrix} 2\lambda_{\rm h} v^2 & \lambda_{\rm hs} v v_{\rm s} \\ \lambda_{\rm hs} v v_{\rm s} & 2\lambda_{\rm s} v_{\rm s}^2 \end{pmatrix},
\ee
with the mixing angle given by
\be
\tan(2\theta) = \frac{v v_{\rm s} \lambda_{\rm hs}}{\lambda_{\rm h}v^2-\lambda_{\rm s}v_{\rm s}^2}.
\ee
In the following analysis we will neglect terms of the order $\mathcal{O}(\lambda_{\rm hs})$ in the mass eigenstates and vacuum expectation values, and therefore work in the limit of zero mixing, unless otherwise noted.

In the non-Abelian model, all three massive vector bosons are degenerate in mass and stable due to a
custodial global $SO(3)$ symmetry of the hidden sector~\cite{Hambye:2008bq}, and in the Abelian case the massive gauge boson is stable due to a remnant $\mathbb{Z}_2$ symmetry, an analogue of the CP symmetry in the visible sector.
Finally, we note that the renormalization group running of couplings is
insignificant up to the Planck scale for the values of couplings we will
consider in the following sections.

\section{Origin of Dark Matter}
\label{abundance}

We consider a scenario where the hidden sector never thermalizes with the SM, and therefore assume that the portal coupling takes a very small value, $\lambda_{\rm hs}\ll1$.
The DM production proceeds as follows: First, an initial abundance of $s$ particles is produced through Higgs decays
~\cite{Chu:2011be}
\be
n_{\rm D}^{\rm initial} \simeq \left. 3\frac{n_{\rm h}^{\rm eq}\Gamma_{h\rightarrow ss}}{H}\right|_{T=m_{\rm h}},
\label{eq:n init}
\ee
where $n_{\rm h}^{\rm eq}$ is the equilibrium number density of Higgs bosons, $H$ is the Hubble rate, and the Higgs decay width into $s$ particles is
\be
\Gamma_{h\rightarrow ss} = \frac{\lambda_{\rm hs}^2 v^2}{32\pi m_{\rm h}} \sqrt{1-\left(\frac{2m_{\rm s}}{m_{\rm h}}\right)^2}.
\label{eq:htoss}
\ee
The yield arises during a short time interval between the moment when the Higgs field acquired a vacuum expectation value around $T\sim m_{\rm h}$, and the moment when the number density of Higgs particles became Boltzmann-suppressed, $T\sim m_{\rm h}/3$. We thus evaluate the above expression at $T\approx m_{\rm h}$.

If the singlet particles were heavy enough, $m_{\rm s} \geq 2m_{\rm A}$, and had no significant interactions within the hidden sector, they would simply decay into the DM particles, $s\rightarrow AA$, resulting in the final yield of twice the abundance of $s$ given by Eq. \eqref{eq:n init}. However, if particle number changing interactions such as $ss\leftrightarrow AA$, $ss\leftrightarrow sss$ and $AA\leftrightarrow AAA$ within the hidden sector are fast, the hidden sector will reach chemical equilibrium
at a temperature $T_{\rm D}\neq T$. Then the final DM abundance is
not given by the usual freeze-in mechanism but by a dark freeze-out \cite{Carlson:1992fn,Chu:2011be,Bernal:2015ova,Bernal:2015xba,Heikinheimo:2016yds,Bernal:2017mqb}, operating in the hidden sector.

Here we examine the scenario where $m_{\rm s} \leq m_{\rm A}$, so that the
hidden sector annihilation process $AA\rightarrow ss$ is kinematically allowed in the non-relativistic limit.
Then, if chemical equilibrium is reached within the hidden sector, the final
abundance is given by the freeze-out of this process, with the freeze-out
temperature approximately set by the condition
$\langle \sigma_{AA\rightarrow ss}v\rangle_{T_{\rm D}}n_A(T_{\rm D}) = H(T)$. Here $n_A$ is the DM number density and $\langle\cdot\rangle_{T_{\rm D}}$ denotes an average over the hidden sector
thermal distribution, with the hidden sector temperature given by
\be
T_{\rm D} =\xi T= \left(\frac{g_*^{\rm SM}\rho_{\rm D}}{g_*^{\rm D}\rho_{\rm SM}}\right)^\frac14 T,
\label{eq:TD}
\ee
where $g_*^{\rm SM(D)}$
denotes the number of relativistic degrees of freedom in the visible 
(hidden) sector, $\rho_{\rm SM}$ and $\rho_{\rm D}$ are the energy densities of the visible and hidden sectors, and we have introduced the notation $\xi=T_{\rm D}/T$ for the ratio of the hidden and visible sector temperatures. The initial value of $\rho_{\rm D}$ is given by $m_{\rm h} n_{\rm D}^{\rm initial}/2$, where the average energy of the DM particles produced from Higgs decays is $m_{\rm h}/2$.

From equations (\ref{eq:n init}) - (\ref{eq:TD}) we see that the hidden sector temperature is controlled by the parameter $\lambda_{\rm hs}$, and vanishes in the decoupling limit $\lambda_{\rm hs}\rightarrow 0$, as this would correspond to the hidden sector not being populated at all. We take the opposing limiting value $\xi=1$ as the limit above which the freeze-in approximation breaks down, {\it i.e.} neglecting the scattering terms from the hidden sector to the SM in the Boltzmann equation for the number density of the $s$ particles is no longer valid, and instead the abundance should be computed assuming kinetic equilibrium between the hidden and visible sectors as in \cite{Escudero:2017yia,Evans:2017kti}. This results in an upper limit for the portal coupling, $\lambda_{\rm hs}\lesssim 6\times 10^{-7}$ in the $SU(2)$ scenario, and $\lambda_{\rm hs}\lesssim 4\times 10^{-7}$ in the $U(1)$ scenario, where the difference originates from the different number of relativistic degrees of freedom in the hidden sector. A similar limit may be obtained by comparing the hidden sector to SM scattering rate to the Hubble rate at the time of DM production, $T\sim m_{\rm h}$.

For the $U(1)$ scenario the relic abundance of DM is approximated as
\be
\Omega_{\rm CDM}h^2 = \frac{1.07\times 10^9\xi x^{\rm FO}\ {\rm GeV}^{-1}}{\sqrt{g_*}M_{\rm P}\langle\sigma_{AA\rightarrow ss} v\rangle},
\label{eq:abundance}
\ee
where $M_{\rm P}$ is the Planck mass and the freeze-out temperature $x^{\rm FO}=m_{\rm A}/T_{\rm D}$ is given by
\be
x^{\rm FO}=\log\left(\xi^2\frac{M_{\rm P} m_{\rm A}\langle\sigma_{AA\rightarrow ss} v\rangle \sqrt{x^{\rm FO}}}{1.66\sqrt{g_*}(2\pi)^\frac32}\right).
\ee
The thermally averaged annihilation cross section is in the nonrelativistic limit given by
\be
\langle \sigma_{AA\rightarrow ss}v\rangle = \frac{9g'^4}{128\pi m_{\rm A}^2}\sqrt{1-\frac{m_{\rm s}^2}{m_{\rm A}^2}}.
\ee

For the $SU(2)$ scenario, in addition to the annihilation process, there is a semi-annihilation channel that is the dominat process away from resonances~\cite{Karam:2015jta}. The thermally averaged annihilation and semi-annihilation cross sections are in the nonrelativistic limit given by~\cite{DiChiara:2015bua}
\bea
\langle\sigma_{AA\rightarrow ss} v\rangle &=& \frac{11m_A^2}{432\pi v_{\rm s}^4}\sqrt{1-\frac{m_{\rm s}^2}{m_{\rm A}^2}}, \\
\langle\sigma_{AA\rightarrow As} v\rangle &=& \frac{m_A^2}{8\pi v_{\rm s}^4}\sqrt{1-\frac{(m_{\rm s}+m_{\rm A})^2}{4m_{\rm A}^2}}.
\eea

The DM abundance is then given by eq. (\ref{eq:abundance}), multiplied by a factor of three to account for the three degenerate DM species, with the replacement $\langle\sigma_{AA\rightarrow ss} v\rangle \rightarrow \langle\sigma_{AA\rightarrow ss} v\rangle+\frac12 \langle\sigma_{AA\rightarrow As} v\rangle$.

After DM has decoupled from the hidden sector heat bath, the remaining massive $s$ particles decay into the SM sector long before Big Bang Nucleosynthesis (BBN) at $T\sim 1$ MeV, and hence do not endanger the success of production of light elements. We have checked that for all values of $m_{\rm s}$ in the region of interest considered in the next section, the lifetime of $s$ remains below one second, corresponding to decay before BBN.

\section{Indirect detection}
\label{indirect_detection}

In the vector DM scenario discussed above, where the DM abundance is determined via dark sector freeze-out of the $AA\rightarrow ss$ annihilation, or the $AA\rightarrow As$ semi-annihilation, followed by $s\rightarrow {\rm SM}$ decays, indirect detection signals from this kind of cascade annihilation process can be expected from regions of high DM density, such as the central region of the Milky Way galaxy, or from DM dominated dwarf spheroidals.

For example, the FERMI-LAT observation of an excess for $\gamma$-rays in the
few GeV energy range from the galactic
center~\cite{Goodenough:2009gk,Hooper:2010mq,Daylan:2014rsa,Calore:2014xka,Escudero:2017yia} has attained considerable attention in the recent years. The
excess was deemed compatible with a DM particle having a mass roughly in the
range $m_{\rm DM}\approx (40-70)\ {\rm GeV}$ and annihilating for example into
$b\bar{b}$ with a velocity averaged cross section \mbox{$\langle\sigma_{\rm
ann}v\rangle \sim 2\times 10^{-26}\ {\rm cm}^3/{\rm s}$}.

The case of a four-body final state resulting from a cascade annihilation, such as in our $U(1)$ scenario, was analyzed in~\cite{Dutta:2015ysa,Profumo:2017obk}. Since we are considering the production of DM originating from Higgs decays, we will constrain our analysis to the mass hierarchy \mbox{$m_{\rm s} \leq m_{\rm A}\lesssim  m_{\rm h}/2$}. The first inequality is needed to allow the $AA\rightarrow ss$ annihilation process, while the second inequality follows from the requirement that the DM particles $A$  
reach chemical equilibrium in the hidden sector after the freeze-in production. Therefore, the region that best fits the excess assuming the $4b$ final state, with $m_A\gtrsim 50\ {\rm GeV}$~\cite{Dutta:2015ysa} lies mostly outside our allowed parameter space. However, as discussed in~\cite{Dutta:2015ysa}, the excess may almost as well be fitted with a $4\tau$ final state, which is the dominant final state in our scenario assuming $2 m_\tau\leq m_s < 2 m_b$.

\begin{figure*}
\begin{center}
\includegraphics[width = 0.49\textwidth]{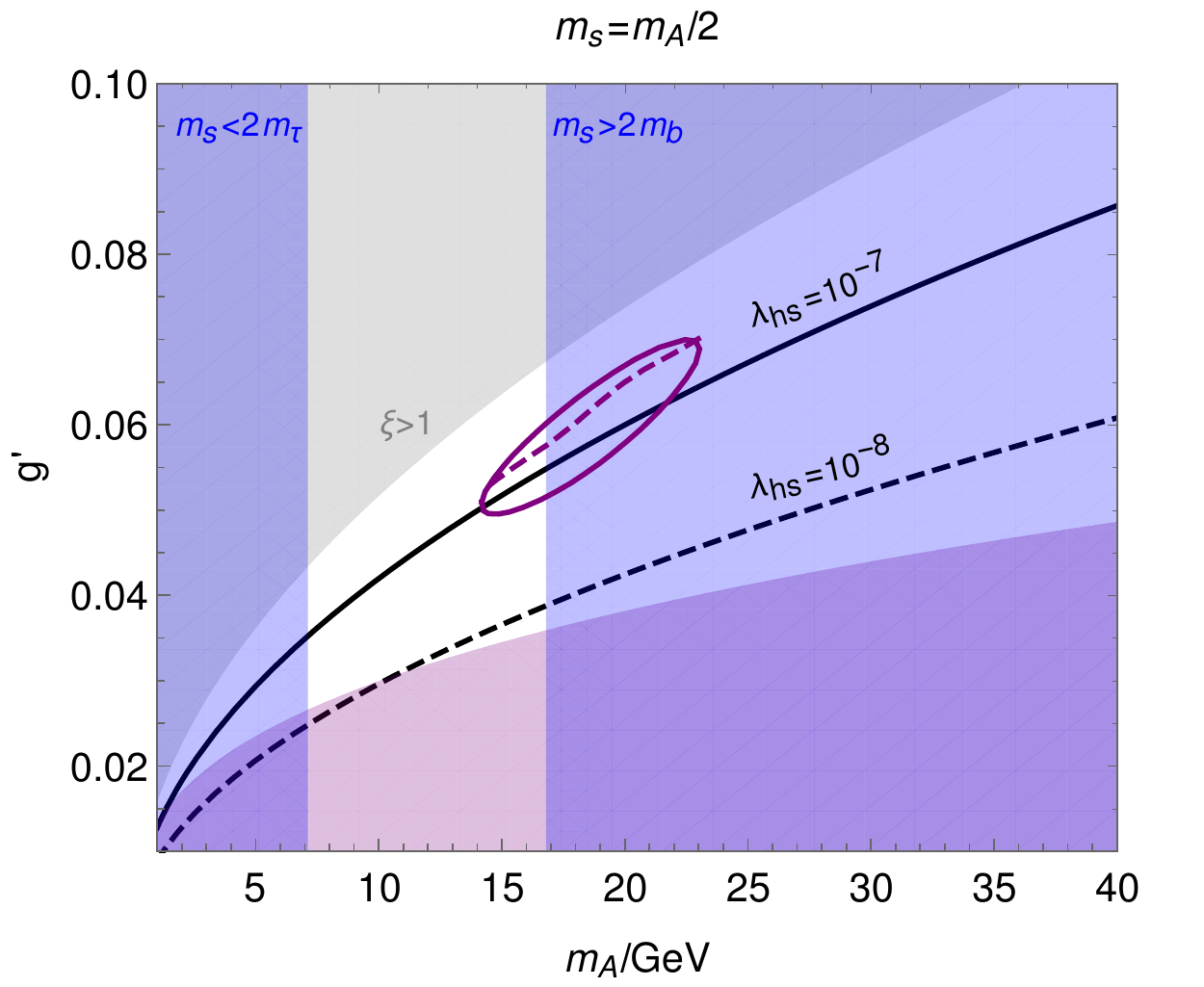}
\includegraphics[width = 0.49\textwidth]{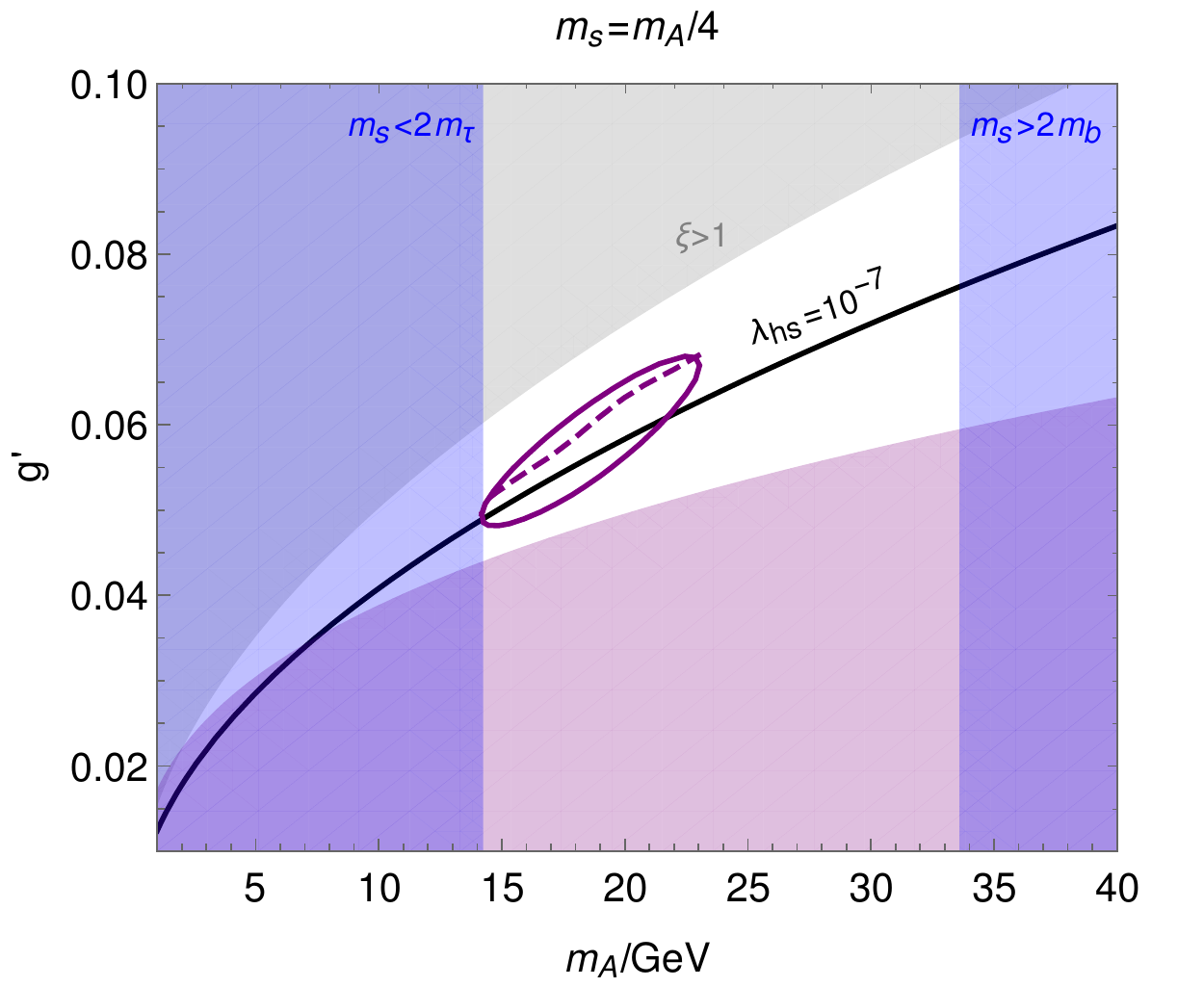}
\caption{The correct vector DM abundance in the $U(1)$-symmetric case is produced along the black solid
($\lambda_{\rm hs}= 10^{-7}$) and dashed ($\lambda_{\rm hs}=10^{-8}$)
 lines via the freeze-out of the $AA\rightarrow ss$ process, following the thermalization of the hidden sector via $ss\rightarrow AA$ and $ss\rightarrow sss$.
The mass hierarchy of the hidden sector is $m_{\rm s}=m_{\rm A}/2$ ($m_{\rm s}=m_{\rm A}/4$) in the left (right) panel.
The region compatible with the GCE according to~\cite{Dutta:2015ysa} is shown
by the purple ellipse, and the upper limit for the cross section from FERMI
dwarf observations is shown by the purple dashed line. The blue shaded regions
show where the decay channel $s\rightarrow\tau\tau$ is not the dominant decay
mode, and therefore can not be used to fit the excess. In the gray shaded region the hidden and visible sector will reach thermal equilibrium, and the standard freeze-out mechanism will determine the DM abundance, and in the purple shaded region the hidden sector will not reach chemical equilibrium, and the DM abundance will be determined by the freeze-in mechanism. Thus the dark freeze-out mechanism is only valid within the white window between the two shaded regions.}
\label{U1plot}
\end{center}
\end{figure*}

In Figure \ref{U1plot} the solid black line shows where the correct DM abundance can be produced via the hidden sector freeze out,
while the purple ellipse shows where the GeV-range Galactic Center Excess (GCE) with the $4\tau$ final state can be simultaneously fitted.
The left and right panels of the figure correspond to two
choices of the hidden sector mass hierarchy: $m_{\rm s}=m_{\rm A}/2$ in the left panel and $m_{\rm s}= m_{\rm A}/4$ in the right.
To estimate the annihilation cross section in the galactic center today, we have used $v=10^{-3}$ as an estimate for the DM velocity dispersion when evaluating the Sommerfeld enhancement factor for the process $AA\rightarrow ss$ \cite{Blum:2016nrz}. 
We have also checked that the scenario is compatible with the upper limit on DM self-interaction cross section, $\sigma_{\rm DM}/m_{\rm DM}\lesssim 1 {\rm cm}^2/{\rm g}$ \cite{Tulin:2017ara}.

There is an intricate interplay with the temperature ratio $\xi$, that is controlled by the value of the portal coupling $\lambda_{\rm hs}$ and the annihilation cross section required to fit both the relic abundance and the GCE today: If the hidden sector temperature was smaller, the corresponding equilibrium density of DM at a given SM temperature would also be smaller, and therefore a smaller annihilation cross section would be needed to produce the observed relic abundance. On the other hand, the photon yield from the galactic center is computed assuming the observed DM abundance, and thus results in a fixed value for the annihilation cross section in order to fit the excess. Therefore, the value of the portal coupling that sets the temperature ratio $\xi$ can be used as a tunable parameter to make these two requirements for the annihilation cross section coincide. This is depicted by the solid and dashed black lines in Figure \ref{U1plot}, showing how the correct DM abundance is produced for two different values of the portal coupling, $\lambda_{\rm hs}=10^{-7}$ and $\lambda_{\rm hs}=10^{-8}$, respectively. This model building tool could be especially useful if future observations would turn out to require a significantly smaller annihilation cross section than what is needed for the correct relic abundance assuming $\xi=1$, i.e. usual freeze-out of DM from a thermal equilibrium between the hidden and visible sectors.

However, in the $U(1)$ model the vector DM particles do not have number changing interactions, such as 
$AA\leftrightarrow AAA$ at the tree level, and therefore the thermalization of the hidden sector must proceed via the $ss\leftrightarrow AA$ and $ss\leftrightarrow sss$ interactions. This results in a lower limit for the scalar self coupling $\lambda_{\rm s}$, below which the hidden sector will not reach internal chemical equilibrium, and the usual freeze-in mechanism will determine the DM abundance. This lower limit is shown in Figure \ref{U1plot} by the purple shaded region, where the dark freeze-out mechanism is not valid. The extend of this region is sensitive to the mass hierarchy within the hidden sector, as the scalar self coupling is determined as a function of the mass ratio and the hidden sector gauge coupling.

\begin{figure*}
\begin{center}
\includegraphics[width = 0.49\textwidth]{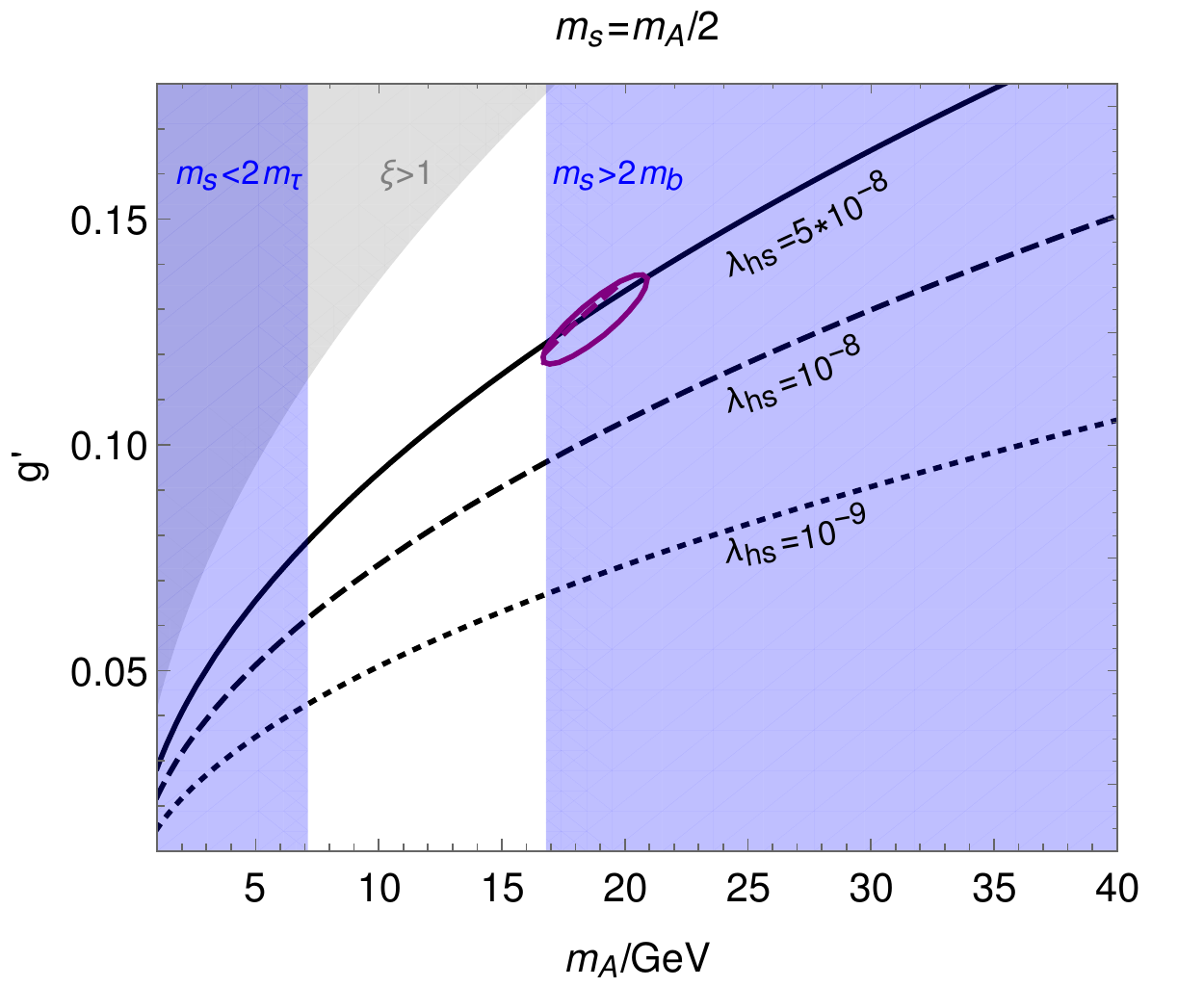}
\includegraphics[width = 0.49\textwidth]{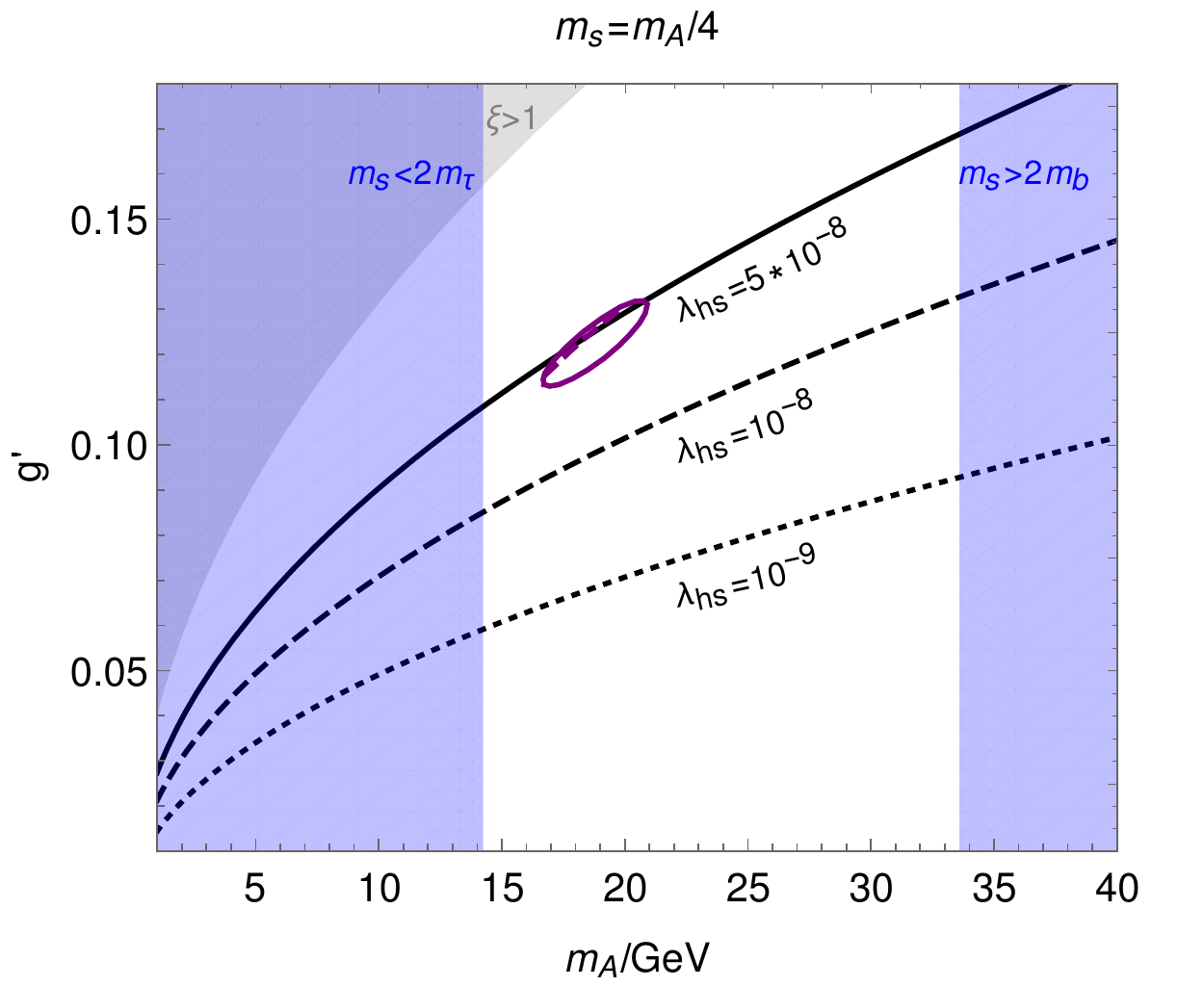}
\caption{Same as Figure \ref{U1plot}, for the $SU(2)$ scenario. Here the DM abundance is dominantly set by the freeze-out of the semi-annihilation process, which is also the source of the indirect detection signal. The correct DM abundance is produced along the black solid ($\lambda_{\rm hs}= 5\times 10^{-8}$), dashed ($\lambda_{\rm hs}=10^{-8}$) and dotted ($\lambda_{\rm hs}=10^{-9}$) lines.}
\label{SU2plot}
\end{center}
\end{figure*}

In Figure \ref{SU2plot} the same information is shown for the $SU(2)$ DM scenario. Here the indirect detection signal results from the semi-annihilation process $AA\rightarrow As$, followed by the decay $s\rightarrow \tau \tau$. Therefore the resulting signature corresponds to that of a usual $\chi\chi\rightarrow \tau \tau$ annihilation process, but with the DM mass scaled by a factor of two, due to one half of the initial state energy escaping back to the hidden sector with the final state $A$ in the semi-annihilation process. We have therefore used the $2\tau$ fit from~\cite{Dutta:2015ysa}, and scaled the DM mass by a factor of two, and the cross section by a factor of four, to account for the reduction of the DM number density due to the scaling of the DM mass. In this scenario the hidden sector reaches internal chemical equilibrium in the whole parameter space shown in the figure, due to the tree level $AA\rightarrow AAA$ process present in the non-Abelian gauge group. Thus the window for tuning the hidden sector temperature is larger in this case.

\section{Conclusions}
\label{conclusions}

We showed that feebly-coupled, frozen-in DM -- which was never in thermal equilibrium with the SM particles 
and is thus of non-thermal origin -- can result in observable indirect detection signals. As benchmark scenarios, we investigated two models where the DM is a spin-1 vector boson of a $U(1)$ or an $SU(2)$ symmetry in a hidden sector connected to the SM via a coupling between a mediator scalar and the SM Higgs boson, $\lambda s^\dagger s\Phi^\dagger\Phi$. We showed that in these models the observed GCE can be fitted simultaneously with
the observed DM abundance.
Consistency with observations requires that the scalar mediator is
lighter than the vector DM candidate and that the ratio of the masses
needs to be within a factor of few from each other, $2\lesssim m_A/m_{\rm s}\lesssim 4$.

Because the hidden sector does not thermalize with the SM, the temperature of the hidden sector equilibrium bath $T_{\rm D}$ is not equal to the SM temperature $T$. This is important, as the temperature ratio $\xi=T_{\rm D}/T$ affects how the DM annihilation cross section is determined by the observed relic abundance.
We showed that attempting to explain the observed GCE within a FIMP model tends to put the model close to the boundary where the model assumptions become inconsistent, i.e. $\lambda_{\rm {hs}}$ grows sufficiently large for the hidden sector to equilibrate with the visible one, $\xi=1$.
However, the features we have uncovered would allow one to tune the expected luminosity of the indirect detection signal. Hence, a weak indirect detection signal, if observed, could be successfully explained in this setup, while it would be difficult within a standard WIMP paradigm where $\xi=1$.

These features are not expected to arise in FIMP models with a fermion DM candidate due to velocity suppression of the relevant cross section. On the other hand, our results are generally applicable in models of FIMP dark matter, where the DM particle is a scalar or vector and couples to a lighter scalar mediator.

\section*{Acknowledgements}
We thank Alexandros Karam for comments. This work has been supported by the Academy of Finland, grants 267842 and 310130. T.T. is supported by the U.K. Science and Technology Facilities Council grant ST/J001546/1.

\bibliography{fimp_detection.bib}

\end{document}